\documentclass[aps,prl,twocolumn,showpacs,floatfix]{revtex4}
\usepackage{epsfig}
\usepackage{graphicx}
\usepackage{dcolumn}
\usepackage{longtable}
\usepackage{amsthm,amsmath}

\begin{document}

\preprint{\today}

\title{Electric dipole moment of $^{129}$Xe atom}

\author{$^1$Yashpal Singh, $^1$B. K. Sahoo{\footnote{bijaya@prl.res.in}} and $^2$B. P. Das}
\affiliation{ $^1$Theoretical Physics Division, Physical Research Laboratory,
Navrangpura, Ahmedabad - 380009, India}
\affiliation{$^2$Theoretical Physics and Astrophysics Group, Indian Institute of Astrophysics, Bangalore-560034, India}

\begin{abstract}
The parity (P) and time-reversal (T) odd coupling constant associated with the
tensor-pseudotensor (T-PT) electron-nucleus interaction and the
nuclear Schiff moment (NSM) have been determined by combining the result of the measurement of the electric dipole moment (EDM) of $^{129}$Xe atom and calculations
based on the relativistic coupled-cluster (RCC) theory. Calculations using various 
relativistic many-body methods have been performed at different levels of 
approximation and their accuracies are estimated
by comparing the results of the calculated dipole polarizability of the ground
state of the above atom with its most precise available experimental data.
The non-linear terms that arise in the 
RCC theory at the singles and doubles
approximation were found to be crucial for achieving high accuracy in the
calculations. Our results for the $^{129}$Xe  EDM due to the odd T-PT interaction and the NSM 
are, respectively, $d_A=0.501 \times 10^{-20} 
C_T \langle \sigma_N \rangle |e|cm$ and $d_A=0.336 \times 
10^{-17} \frac{S}{|e|~fm^3} |e| cm$. These results in combination with the 
future EDM measurements in atomic Xe  could provide the most accurate limits for
the T-PT coupling constant and NSM.
\end{abstract}


\maketitle
The search for the electric dipole moment (EDM) is now in its seventh decade
\cite{khriplovich,roberts}. The observation of an EDM of an elementary particle or a composite system
would be an unambiguous signature  of the violations of parity (P) 
and time-reversal (T) symmetries. T violation implies charge conjugation-parity (CP) 
violation via the CPT theorem \cite{luders}. The standard model (SM) of elementary particle physics provides explanations of the 
experimentally observed hadronic CP violation in the decays of neutral K \cite{christenson} and B \cite{abe,aubert,aaij} mesons, but the amount of CP violation predicted by the SM is not sufficient to account for the matter-antimatter asymmetry
in the Universe \cite{dine}.
The current limits for CP violating coupling constants deduced
from the atomic EDMs are several orders of magnitude higher than the predictions of these quantities by the SM \cite{pospelov,barr,ramsey}. 
In addition, atomic EDMs can probe CP violation originating from leptonic, semi-leptonic and hadronic CP sources.
Combining atomic EDM measurements with high precision many-body calculations, it is possible to obtain various CP violating coupling
constants at the levels of the nucleus and the electron. Newly proposed EDM experiments on diamagnetic and paramagnetic atoms hold the promise of 
improving the sensitivity of the current measurements by at least a few orders of magnitude \cite{furukawa-xe,inoue-xe,rand-rn,weiss,heinzen}. 
The EDMs of diamagnetic atoms arise predominantly from the electron-nucleus tensor-pseudotensor (T-PT) interaction and 
interaction of electrons with the nuclear Schiff moment (NSM) \cite{barr1}. The electron-nucleus T-PT interaction is due to the CP violating
electron-nucleon interactions which translates to CP violating electron-quark interactions at the level of elementary particles. The NSM on
the other hand could exist due to CP violating nucleon-nucleon interactions and the EDM of nucleons and both of them in turn could originate from 
CP violating quark-quark interactions or EDMs and chromo EDMs of quarks. In order to obtain precise limits for the coupling constants of these interactions 
and EDMs of quarks, it is necessary to perform both experiments and calculations as accurately as possible on suitable atoms.

 To date the best limit for a diamagnetic atomic EDM is obtained from $^{199}$Hg atom as
$d_A < 3.1 \times 10^{-29} \ |e| cm$ \cite{griffith} and the next best limit comes from an
earlier measurement on $^{129}$Xe atom as $d_A < 4.1 \times 10^{-27} \ |e| cm$ \cite{rosenberry}. Both 
$^{129}$Xe and $^{199}$Hg isotopes are good choices for carrying out EDM 
measurements as they have nuclear spin $I=1/2$ and therefore the interaction with the octupole moment vanishes. 
Owing to the fact that the matrix elements of the T-PT and NSM interaction
Hamiltonians increase with the size of the atomic system, their enhancements in atomic Hg are larger than Xe. 
However, the new proposals on EDM measurements in $^{129}$Xe argue in favor of carrying out the experiment in
this isotope because of its larger spin relaxation time \cite{inoue-xe}.
As a matter of fact, three research groups around the world are now
actively involved in Xe EDM experiments \cite{inoue-xe, fierlinger, schmidt}.
Inoue {\it et al.} have proposed to utilize the nuclear spin maser
technique \cite{yoshimi} to surpass the limit provided by the Hg EDM 
measurement.

In this Letter, we report the results of our systematic theoretical studies of the
P- and T- odd coupling constant for the T-PT interaction and of the NSM in $^{129}$Xe. 
To this end, we have developed many-body methods in the framework of the
third order many-body perturbation theory (MBPT(3))  for a better understanding of the
different classes of correlation effects, the coupled-perturbed-Hartree-Fock (CPHF) method in order to reproduce the
previously reported results and the relativistic coupled-cluster (RCC) theory to bring to light the roles of both the CPHF and non-CPHF contributions (e.g. 
pair-correlation effects) to all orders in the residual Coulomb interaction (difference
between the exact two-body Coulomb and the mean-field interactions). 
In the present work, we consider only one hole-one particle and two hole-two particle excitations, i.e. the CCSD method and its linearized
approximation, the LCCSD method. The ground state of a closed shell atom like Xe can be exactly described in the RCC theory by
\begin{eqnarray}
 | \Psi \rangle &=& e^T |\Phi_0 \rangle,
\label{eqcc}
\end{eqnarray}
where the cluster operator $T$ generates single and double excitations from the Dirac-Hartree-Fock
(DF) wave function $|\Phi_0 \rangle$ by defining $T=T_1+T_2$. These operators can be expressed
in second quantization notation using hole and particle creation and annihilation operators as
\begin{eqnarray}
T_1=\sum_{a,p} a_p^{\dagger} a_a t_a^p ~~ \mbox{and}~~ T_2= \frac{1}{4} \sum_{a,b,p,q} a_p^{\dagger}a_q^{\dagger} a_b a_a t_{ab}^{pq}
\label{3}
\end{eqnarray} 
with $t_a^p$ and $t_{ab}^{pq}$ are the excitation amplitudes from the occupied 
orbitals denoted by $a,b$ to the unoccupied orbitals denoted by $p,q$ which embody
correlation effects among the electrons to all orders.
\begin{table}[t]
\caption{\label{tab1} Results of $\alpha$ in $e a_0^3$, $\overline{\eta}=10^{20} \times \eta$ and $\overline{\zeta}=10^{17} \times \zeta$ for the ground state of Xe
using different many-body methods. The CCSD results given in bold fonts are the recommended values from the calculations on the physical ground.}
\begin{ruledtabular}
\begin{tabular}{lccccccc}
Method of & \multicolumn{3}{c}{\textrm{This work}}&\multicolumn{4}{c}{\textrm{Others}}\\
  \cline{2-4}  \cline{5-8}\\
 Evaluation  & \textrm{$\alpha$} & $\overline{\eta}$& $\overline{\zeta}$ & $\alpha$ & $\overline{\eta}$ &$\overline{\zeta}$ & Ref. \\
\hline        \\
DF & 26.918&0.447 &0.288  &  & 0.45 & 0.29 & \cite{dzuba} \\
MBPT(2)& 23.388&0.405 &0.266  &                                \\ 
MBPT(3)&18.693 &0.515 &0.339  & & 0.52 &      & \cite{martensson} \\
CPHF   &26.987 &0.562 &0.375  &  & 0.57 & 0.38 & \cite{dzuba}\\
       &       &      &       & 27.7 & 0.564  &  & \cite{latha} \\
LCCSD  &27.484 &0.608 &0.417  &                      &  \\
CCSD   & {\bf 27.744} & {\bf 0.501} & {\bf 0.336}  &            & \\
Experiment & \multicolumn{2}{l}{27.815(27)} &  &  &  &  & \cite{hohm} \\
\end{tabular}                    
\end{ruledtabular}   
\end{table}  
We consider the Dirac-Coulomb (DC) Hamiltonian which in atomic unit (au) is given by
\begin{eqnarray}
H &=&\sum_i [ c\mbox{\boldmath$\alpha_D$}\cdot \textbf{p}_i+(\beta_D -1)c^2+
V_{n}(r_i)] + \sum_{i,j>i} \frac{1}{r_{ij}}, \ \ \ \ \
\label{eq1}
\end{eqnarray}
where $c$ is the velocity of light in vacuum, $\alpha_D$ and $\beta_D$ are the 
Dirac matrices, $V_n$ denotes the nuclear potential obtained using the Fermi-charge 
distribution and $\frac{1}{r_{ij}}$ is
the dominant inter-electronic Coulombic repulsion. We also take into account one 
order of an additional operator $H_{add}$ which is either the dipole operator $D$ for the
evaluation of dipole polarizability ($\alpha$) or the P- and T- violating interaction Hamiltonians
for determining their corresponding couplings coefficients. The T-PT and the
NSM interaction Hamiltonians are given by \cite{dzuba}
 \begin{eqnarray}
  H_{EDM}^{TPT}=\frac{iG_FC_T}{\sqrt{2}} \sum \mbox{\boldmath $\sigma_n \cdot \gamma_D$} \rho_n(r)
 \end{eqnarray}
and
 \begin{eqnarray}
  H_{EDM}^{NSM}= \frac{3{\bf S.r}}{B_4} \rho_n(r),
 \end{eqnarray}
respectively, with $G_F$ is the Fermi coupling constant, $C_T$ is the T-PT coupling constant, 
{\boldmath$\sigma_n$}$= \langle \sigma_n \rangle \frac{{\bf I}}{I}$ is the Pauli spinor of the nucleus for the nuclear spin $I$, {\boldmath$\gamma_D$} represents the Dirac
matrices, $\rho_n(r)$ is the nuclear density, 
${\bf S}=S \frac{{\bf I}}{I}$ is the NSM and $B_4=\int_0^{\infty} dr r^4 \rho_n(r)$.

To distinguish between the correlations only due to the Coulomb and the combined Coulomb and the
additional interaction, we further define
\begin{eqnarray}
T=T^{(0)}+T^{(1)} 
\label{eqn4}
\end{eqnarray}
for the cluster operators $T^{(0)}$ and $T^{(1)}$ that account for the correlations only due to the
Coulomb interaction and the combined Coulomb-additional interactions 
respectively. To ensure the inclusion of only one order of the additional 
interaction in the wave function, we express
\begin{eqnarray}
|\Psi \rangle &\simeq& \left ( e^{T^{(0)}}+e^{T^{(0)}} T^{(1)} \right )|\Phi_0 \rangle \nonumber \\
             &=& |\Psi^{(0)} \rangle + |\Psi^{(1)} \rangle ,
\label{eq4}
\end{eqnarray} 
where $|\Psi^{(0)} \rangle$ and $|\Psi^{(1)} \rangle$ are the
unperturbed and the first order perturbed wave functions due to the additional 
interaction.
Owing to the nature of the additional operators, the first order perturbed wave function is
an admixture of both the even and odd parities. The working equations for evaluating the 
excitation amplitudes of these RCC operators are described in \cite{yashpal}.

Using the generalized Bloch equation, we can also express \cite{yashpal}
\begin{eqnarray}
|\Psi\rangle &=& \Omega^{(0)} |\Phi_0 \rangle + \Omega^{(1)} |\Phi_0 \rangle \nonumber \\
  &=& \sum_{k} [\Omega^{(k, 0)} + \Omega^{(k,1)} ] |\Phi_0 \rangle, 
\end{eqnarray} 
where the $\Omega$s are known as the wave operators with $\Omega^{(0, 0)}=1$ and $\Omega^{(1, 0)}=H_{add}$ and $k$ represents the order of interactions due to the Coulomb repulsion.
In the MBPT(3) method, we restrict $k$ up to 2. The diagrams that make important contributions in this approximation are given explicitly in \cite{yashpal}.
\begin{figure}[t]
\includegraphics[width=8cm, height=3cm, clip=true]{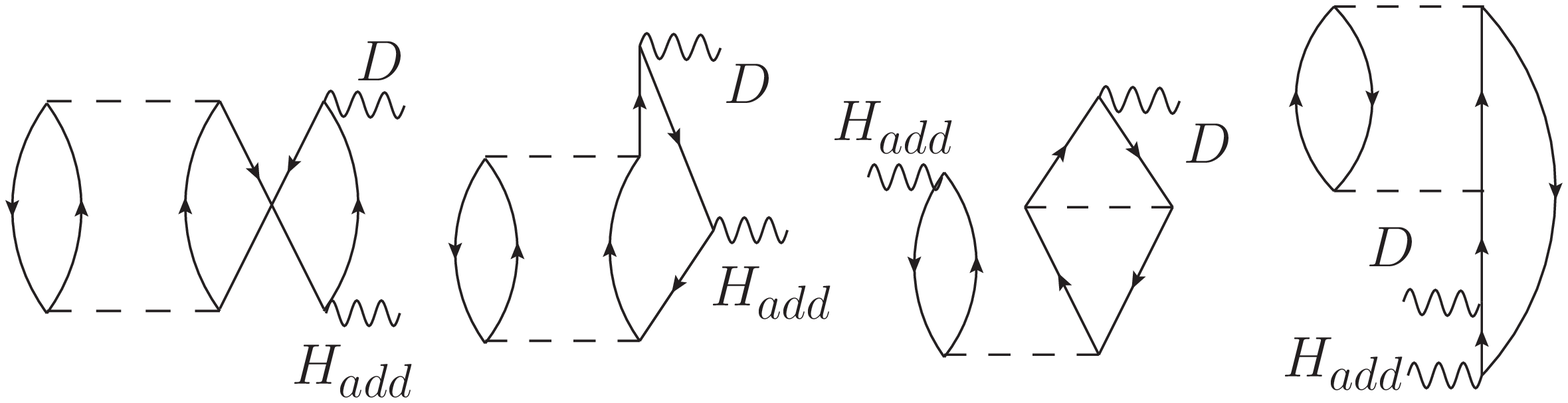}
\caption{Example of few dominant non-CPHF diagrams from the MBPT(3) method involving
$D$ and the corresponding perturbed interaction operator $H_{add}$.}
\label{fig1}
\end{figure}
In the CPHF method, we consider $\Omega^{(k,0)} \approx \Omega^{(0,0)}$ and $\Omega^{(k,1)}$ is evaluated to 
infinite order by restricting it only to one hole-one particle excitations by defining
\begin{eqnarray}
\Omega_{a \rightarrow p}^{(\infty, 1)} &=&  \sum_{k=1}^{\infty} \sum_{b,q} {\Huge \{} \frac{[\langle pb | \frac{1}{r_{ij}} | aq \rangle 
- \langle pb | \frac{1}{r_{ij}} | qa \rangle] \Omega_{b \rightarrow q}^{(k-1,1)} } {\epsilon_a - \epsilon_p}  \nonumber \\ 
&& + \frac{ \Omega_{b \rightarrow q}^{{(k-1,1)}^{\dagger}}[\langle pq | \frac{1}{r_{ij}} | ab \rangle - \langle pq | \frac{1}{r_{ij}} | ba \rangle] 
}{\epsilon_a-\epsilon_p} {\Huge \}},
\end{eqnarray} 
with $ \Omega_{a\rightarrow p}^{(0,1)}= - \frac{\langle p|H_{add}|a\rangle}{\epsilon_p -\epsilon_a}$,
$\epsilon$'s are the orbital energies and $a \rightarrow p$ represents
single excitations from $|\Phi_0 \rangle$ by replacing one of its occupied orbitals $a$ by a virtual orbital $p$.

\begin{table}[t]
\caption{\label{tab2} Explicit contributions to the $\alpha$ in $e a_0^3$, $\overline{\eta}=10^{20} \times \eta$ and
$\overline{\zeta}=10^{17} \times \zeta$ values from various CCSD terms.}
\begin{ruledtabular}
\begin{tabular}{lccc}
\textrm{Term}&$\alpha $& $\overline{\eta}$ & $\overline{\zeta}$ \\
\hline        \\
$\overline{D}T_1^{(1)} +c.c$                &  26.246  &0.506    & 0.338   \\
$T_1^{(0)\dagger}\overline{D}T_2^{(1)} +c.c$&   0.008  &$\sim$0  & $\sim$0 \\
$T_2^{(0)\dagger}\overline{D}T_2^{(1)} +c.c$&   1.395  &$-$0.005 & $-$0.001\\
$Extra$                                       &   0.095  &$\sim$0  & $-$0.001\\
\end{tabular} 
\end{ruledtabular}   
\end{table}                             
Using the many-body tools discussed above, we evaluate $X$ representing polarizability
 $\alpha$, $\eta=\frac{d_A}{\langle \sigma_N \rangle C_T}$ or $\zeta=\frac{d_A}{S/(|e|~fm^3)}$ 
by considering the appropriate additional operator using the general expression
\begin{eqnarray}
 X&=& 2 \frac{\langle \Psi^{(0)}|D|\Psi^{(1)}\rangle}{\langle \Psi^{(0)}|\Psi^{(0)} \rangle} .
\end{eqnarray}
In the MBPT(3) method, we have
\begin{eqnarray}
 X&=& 2 \frac{ \sum_{k=0}^{m=k+1,2} \langle \Phi_0| \Omega^{{(m-k-1,0)}^{\dagger}} D \Omega^{(k,1)} |\Phi_0 \rangle}{\sum_{k=0}^{m=k+1,2} \langle \Phi_0| \Omega^{{(m-k-1,0)}^{\dagger}} \Omega^{(k,0)} |\Phi_0 \rangle}.
\end{eqnarray}
Therefore, the lowest order MBPT(1) with $k=0$ corresponds to the DF approximation
and the intermediate MBPT(2) approximation follows with $k=1$.

The above expression yields the forms
 $X= 2 \langle \Phi_0| \{D \Omega^{(\infty,1)} \}_{con} |\Phi_0 \rangle$ 
in the CPHF method and
$X=2\langle\Phi_0 | \{\overbrace{D}T^{(1)} \}_{con}|\Phi_0 \rangle$ in the RCC
theory with $\overbrace{D} = (1+T^{{(0)}^{\dagger}}) D$ in the LCCSD method and
$\overbrace{D} = e^{{T^{(0)}}^{\dagger}}De^{T^{(0)}}$ is a non-truncating series
in the CCSD method. The subscript $con$ implies that all the terms inside the
curly bracket are connected. We have described in an earlier work the procedure for
evaluating the diagrams that make the dominant contributions to  $\overbrace{D}$ 
\cite{yashpal}.

\begin{figure}[t]
\includegraphics[width=3cm,height=2.5cm, clip=true]{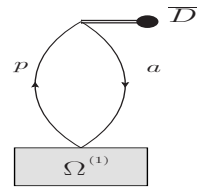}
\caption{Diagram involving effective one-body dipole operator $\overline{D}$ 
and the perturbed wave operator $\Omega^{(1)}$ that accounts for the contributions from
 the singly excited configurations.}
\label{fig2}
\end{figure}
We calculate $\alpha$ for
the ground state of Xe by the methods mentioned above to test their accuracies. The most precise 
measured value of this quantity is
reported as $27.815(27)$ $e a_0^3$ \cite{hohm}. In Table \ref{tab1}, we present
the calculated $\alpha$, $\eta$ and $\zeta$ values along with the experimental
and previously reported results. As can be seen from this table the DF result for 
$\alpha$ is close to the experimental result, but this is not the case when correlation effects
are added via the MBPT(2) and MBPT(3) methods. The results of the all order CPHF, LCCSD and CCSD methods
are in good agreement with the measured value, but the CCSD result is more accurate than the former two methods.
The rationale for considering the non-linear RCC terms in the singles and doubles approximation
for the precise evaluation of the ground state properties of Xe atom is therefore justified. It is also significant to note that the
EDM  enhancement factors exhibit different correlation trends than those of polarizability. The results increase gradually 
from the DF level after the inclusion 
of the correlation effects in the passage from the MBPT to LCCSD, and after that they decrease at the 
CCSD level. 
With reference to the $\alpha$ calculations, the CCSD results, which are marked in bold fonts in the above 
table, are clearly the most accurate. This is evident on physical grounds.

\begin{table*}[t]
\caption{\label{tab3} Contributions from various matrix elements and from various angular momentum symmetry
groups at the DF, lowest order CPHF (denoted by MBPT($l$-CPHF)), CPHF and CCSD methods
to the $\alpha$ in $e a_0^3$, $\overline{\eta}=10^{20} \times \eta$ and
$\overline{\zeta}=10^{17} \times \zeta$ values. 
Here the summation indices $n$ and $m$ represent for the occupied and unoccupied orbitals, respectively.}
\begin{ruledtabular}
\begin{tabular}{lcccccccccccccc}
 Excitation(s) & \multicolumn{3}{c}{\textrm{DF}} & \multicolumn{3}{c}{\textrm{MBPT($l$-CPHF})} & \multicolumn{3}{c}{\textrm{CPHF}}& \multicolumn{3}{c}{\textrm{CCSD}} \\
           \cline{2-4} \cline{5-7}  \cline{8-10}  \cline{11-13}\\
 ($ a\rightarrow p$) & $\alpha$  & $\overline{\eta}$  & $\overline{\zeta}$  & $\alpha$ & $\overline{\eta}$  &  $\overline{\zeta}$      & $\alpha$ & $\overline{\eta}$ & $\overline{\zeta}$  & $\alpha$& $\overline{\eta}$ & $\overline{\zeta}$  \\
\hline        \\
 $5p_{1/2}-7s$ & 0.248       & 0.030 &0.007  & 0.336    &  0.056  &  0.016   & 0.380    &  0.062 & 0.016  & 0.352  & 0.050 &  0.014   \\
 $5p_{1/2}-8s$ & 0.517       & 0.090 &0.022  & 0.690    &  0.159  &  0.045   & 0.769    &  0.172 & 0.045  & 0.733  & 0.145 &  0.039   \\
 $5p_{1/2}-9s$ & 0.237       & 0.106 &0.025  & 0.284    &  0.166  &  0.044   & 0.301    &  0.174 & 0.044  & 0.309  & 0.157 &  0.041   \\
 $5p_{3/2}-7s$ & 0.844       &$\sim$0&0.015  &  1.136   &  0.005  &  0.036   &  1.314   & 0.007  & 0.036  &  1.202 & 0.001 &  0.031   \\
 $5p_{3/2}-8s$ & 1.558       &$\sim$0&0.043  &  2.056   &  0.014  &  0.093   &  2.351   & 0.018  & 0.093  &  2.261 & 0.024 &  0.082   \\
 $5p_{3/2}-9s$ & 0.583       &$\sim$0&0.044  &  0.678   &  0.012  &  0.081   &  0.745   & 0.015  & 0.081  &  0.809 & 0.017 &  0.076   \\
 $5p_{1/2}-7d_{3/2}$ &2.267  &$\sim$0&$\sim$0& 2.200   &$-$0.003 &  $-$0.008 &  2.407   &$-$0.006&$-$0.008&  2.259 &$-$0.011&$-$0.008 \\
 $5p_{1/2}-8d_{3/2}$ &3.454  &$\sim$0&$\sim$0& 2.595    &$-$0.013 &$-$0.020  &  2.882   &$-$0.022&$-$0.020&  2.925 &$-$0.028&$-$0.018 \\
 $5p_{3/2}-7d_{5/2}$ &5.667  &$\sim$0&$\sim$0& 5.747    &$-$0.027 & $-$0.018 &  6.365   &$-$0.039&$-$0.018&  5.827 &$-$0.031&$-$0.018 \\
 $5p_{3/2}-8d_{5/2}$ &7.054  &$\sim$0&$\sim$0& 5.749    &$-$0.048 &$-$0.037  &  6.267   &$-$0.071&$-$0.037&  6.207 &$-$0.057&$-$0.035 \\
 & & & \\
 $\sum_{n,m} (ns - mp_{1/2})$       & 0.013  & 0.121   & 0.029    &0.049    &0.142     &0.036     &0.046   &0.144     & 0.036     &0.046  &0.152    &0.038   \\
 $\sum_{n,m} (ns - mp_{3/2})$       & 0.010  & $\sim$0 & 0.036    &0.025    &0.003     &0.042     &0.018   &0.003     & 0.042     &0.037  &0.004    &0.048   \\
 $\sum_{n,m} (np_{1/2}-ms)$       & 1.064  & 0.326   & 0.078    &1.382    &0.500     &0.136     &1.532   &0.529     & 0.136     &1.474  &0.466    &0.122   \\
 $\sum_{n,m} (np_{3/2}-ms)$       & 3.183  & $\sim$0 & 0.144    &4.111    &0.036     &0.265     &4.696   &0.046     & 0.265     &4.536  &0.057    &0.241   \\
 $\sum_{n,m} (np_{1/2} - md_{3/2})$ & 6.293  & $\sim$0 & $-$0.001 &4.993    &$-$0.022  &$-$0.033  &5.582   &$-$0.038  & $-$0.033  &5.539  &$-$0.047 &$-$0.031\\
 $\sum_{n,m} (np_{3/2} - md_{3/2})$ & 1.545  & $\sim$0 & $\sim$0  &1.326    &$-$0.003  &$-$0.006  &1.501   &   0.003  & $-$0.006  &1.375  &$-$0.006 &$-$0.007\\
 $\sum_{n,m} (np_{3/2} - md_{5/2})$ & 13.860 & $\sim$0 & $\sim$0  &11.887   &$-$0.082  &$-$0.064  &13.428  &$-$0.125  & $-$0.064  &12.871 &$-$0.099 &$-$0.060\\
\end{tabular}
\end{ruledtabular}   
\end{table*}                       
The results of calculations by others for $\alpha$, $\eta$ and $\zeta$ \cite{martensson,
dzuba,latha} as well as the methods used to calculate them are also given in Table \ref{tab1}.
As can be seen in that table, we have successfully reproduced the results of the previous calculations at the same level of
approximation and we have gone beyond these approximations for obtaining accurate results. We present our results performing the calculations using the MBPT(3), LCCSD 
and CCSD methods in Table \ref{tab1}. These results provide useful insights into the role
of different types of correlation effects. From the MBPT(3) calculations, we find that 
certain non-CPHF type diagrams, for example the diagrams shown in Fig. \ref{fig1},
contribute substantially with opposite signs to those of the DF values in all the
above quantities leading to large cancellations in the final results. Indeed this
is the main reason why the CPHF method over estimates the EDM enhancement 
factors compared to the CCSD method. In fact many of these MBPT(3) diagrams correspond
to the non-linear terms of the CCSD method, hence their contributions are
absent in the LCCSD method. Therefore, the LCCSD method also over 
estimates these results even though they account for some of the lower
order non-CPHF contributions.

We present the contributions from the individual CCSD terms in Table \ref{tab2} to highlight
the importance of various correlation effects. It can be seen in this table that by far the most important
contributions comes from $\overline{D}T_1^{(1)}$ term followed by 
$T_2^{{(0)}^{\dagger}} \overline{D}T_2^{(1)}$, where $\overline{D}$ is the effective
one-body term of $\overbrace{D}$ and the contributions from the other terms are almost negligible.
To carry out an analysis similar to the one given in \cite{latha}, we 
find the contributions from various orbitals that correspond to various singly excited intermediate configurations for different properties which are
given in Table \ref{tab3}. These results are evaluated using the diagram shown in
Fig. \ref{fig2} with the corresponding $\Omega^{(1)}$ operator from the
the DF, MBPT(2) containing diagrams that correspond
only to the lowest order CPHF (denoted by MBPT($l$-CPHF)), CPHF and
CCSD methods. We also present the sum of contributions from the orbitals belonging to a particular
category of angular momentum excitations to demonstrate their importance in 
obtaining the properties that have been calculated.  The information provided in all the three tables together clearly expound
the reasons for the different trends in the correlation effects in the calculations of
$\alpha$, $\eta$ and $\zeta$.

By combining our CCSD results for $\eta$ and $\zeta$ with the available experimental limit for $^{129}$Xe EDM, 
$d_a(^{129} \text{Xe}) < 4.1 \times 10^{-27} |e|cm$, we get the
limits $C_T<1.6\times 10^{-6}$ and $S<1.2\times 10^{-9}~ |e| fm^3$. These are not superior to the limits extracted from $^{199}$Hg \cite{lathalett, dzuba}, which are about three 
orders of magnitude lower. However, the experiments on 
$^{129}$Xe \cite{inoue-xe,fierlinger,schmidt} that are underway have the potential to improve the current sensitivity by about three to four orders of magnitude.
It therefore seems very likely that the best limits for both $C_T$ and $S$ could be obtained by combining our calculated values presented in this work and the results of the new
generation of experiments for $^{129}$Xe when they come to fruition. This limit
for $S$ in conjunction with the recent nuclear structure calculations 
\cite{yoshinaga} and quantum chromodynamics (QCD), would yield new 
limits for $\theta_{QCD}$ and CP violating coupling constants involving chromo 
EDMs of quarks.

We acknowledge useful discussions with Professor K. Asahi. This work was supported partly by INSA-JSPS under project
no. IA/INSA-JSPS Project/2013-2016/February 28,2013/4098. The computations were carried out using the 3TFLOP HPC cluster
at Physical Research Laboratory, Ahmedabad.


\begin{thebibliography}{10}
\bibitem{khriplovich}
I. B. Khriplovich and S. K. Lamoreaux, {\it CP violation without strangeness. Electric dipole moments of particles, atoms, and molecules}, (Springer, Berlin, 1997).
\bibitem{roberts}
B. L. Roberts and W. J. Marciano, {\it Lepton Dipole Moments, Advanced series on Directions in High Energy Physics}, vol. 20, World Scientific, Singapore (2010).
\bibitem{luders}
G. Luders, Ann. Phys. (N.Y.) {\bf 281}, 1004 (2000).
\bibitem{christenson}
J. H. Christenson, J. W. Cronin, V. L. Fitch, and R. Turlay, Phys. Rev. Lett. {\bf 13}, 138 (1964).
\bibitem{abe}
K. Abe {\it et al}., Phys. Rev. Lett. {\bf 87}, 091802 (2001).
\bibitem{aubert}
B. Aubert {\it et al}., Phys. Rev. Lett. {\bf 87}, 091801 (2001).
\bibitem{aaij}
R. Aaij {\it et al}., Phys. Rev. Lett. {\bf 110}, 221601 (2013).
\bibitem{dine}
M. Dine and A. Kusenko, Rev. Mod. Phys. {\bf 76}, 1 (2003).
\bibitem{pospelov}
M. Pospelov and A. Ritz, Ann. Phys. (N.Y.) {\bf 318}, 119 (2005).
\bibitem{barr}
S. M. Barr, Int. J. Mod. Phys. A {\bf 8}, 209 (1993).
\bibitem{ramsey}
M. J. Ramsey-Musolf and S. Su, Phys. Rep. {\bf 456}, 1 (2008).
\bibitem{furukawa-xe}
T. Furukawa {\it et al}., J. Phys. Conf. Ser. {\bf 312}, 102005 (2011).
\bibitem{inoue-xe}
T. Inoue {\it et al}., Hyperfine Interactions (Springer Netherlands) {\bf 220}, 59 (2013).
\bibitem{rand-rn}
E. T. Rand {\it et al}., J. Phys. Conf. Ser. {\bf 312}, 102013 (2011).
\bibitem{weiss}
D. S. Weiss, Private communication.
\bibitem{heinzen}
D. Heinzen, Private communication. 
\bibitem{barr1}
S. M. Barr, Phys. Rev. D {\bf 45}, 4148 (1992).
\bibitem{griffith}
W. C. Griffith {\it et al}., Phys. Rev. Lett. {\bf 102}, 101601 (2009).
\bibitem{rosenberry}
M. A. Rosenberry and T. E. Chupp, Phys. Rev. Lett. {\bf 86}, 22 (2001).
\bibitem{fierlinger}
P. Fierlinger {\it et al}., {\it Cluster of Excellence for Fundamental Physics}, Technische Universit\"{a}t M\"{u}nchen
({\it http://www.universe-cluster.de/fierlinger/xedm.html}).
\bibitem{schmidt}
U. Schmidt {\it et al}. {\it Collaboration of the Helium Xenon EDM Experiment}, Physikalisches Institut, University of Heidelberg 
({\it http://www.physi.uni-heidelberg.de/Forschung/ANP/XenonEDM/Team}).
\bibitem{yoshimi}
A. Yoshimi {\it et al}., Phys. Lett. A {\bf 304}, 13 (2002).
\bibitem{dzuba}
V. A. Dzuba, V. V. Flambaum and S.G. Porsev, Phys. Rev. A {\bf 80}, 032120 (2009).
\bibitem{martensson}
A. M. M\r{a}rtensson-Pendrill, Phys. Rev. Lett. {\bf 54}, 1153 (1985).
\bibitem{latha}
K. V. P. Latha and P. R. Amjith, Phys. Rev. A {\bf 87}, 022509 (2013).
\bibitem{hohm}
U. Hohm and K. Kerl, Mol. Phys. {\bf 69}, 819 (1990).
\bibitem{yashpal}
Y. Singh, B. K. Sahoo and B. P. Das, Phys. Rev. A {\bf  88}, 062504 (2013).
\bibitem{lathalett}
K. V. P. Latha, D. Angom, B. P. Das and D. Mukherjee, Phys. Rev. Lett. {\bf 103}, 083001 (2009).
\bibitem{yoshinaga} N. Yoshinaga, K. Higashiyama, R. Arai and E. Teruya, Phys. Rev. C {\bf 87}, 044332 (2013).
\end{thebibliography}
\end{document}